# Optimizing RAG Techniques for Automotive Industry PDF Chatbots: A Case Study with Locally Deployed Ollama Models

Fei Liu *

China Automotive Technology & Research Center, liufei@catarc.ac.cn

Zejun Kang

China Automotive Technology & Research Center, kangzejun@catarc.ac.cn

Xing Han

China Automotive Technology & Research Center, hanxing@catarc.ac.cn

With the growing demand for offline PDF chatbots in automotive industrial production environments, optimizing the deployment of large language models (LLMs) in local, low-performance settings has become increasingly important. This study focuses on enhancing Retrieval-Augmented Generation (RAG) techniques for processing complex automotive industry documents using locally deployed Ollama models.

Based on the Langchain framework, we propose a multi-dimensional optimization approach for Ollama's local RAG implementation. Our method addresses key challenges in automotive document processing, including multi-column layouts and technical specifications. We introduce improvements in PDF processing, retrieval mechanisms, and context compression, tailored to the unique characteristics of automotive industry documents. Additionally, we design custom classes supporting embedding pipelines and an agent supporting self-RAG based on LangGraph best practices.

To evaluate our approach, we constructed a proprietary dataset comprising typical automotive industry documents, including technical reports and corporate regulations. We compared our optimized RAG model and self-RAG agent against a naive RAG baseline across three datasets: our automotive industry dataset, QReCC, and CoQA. Results demonstrate significant improvements in context precision, context recall, answer relevancy, and faithfulness, with particularly notable performance on the automotive industry dataset.

Our optimization scheme provides an effective solution for deploying local RAG systems in the automotive sector, addressing the specific needs of PDF chatbots in industrial production environments. This research has important implications for advancing information processing and intelligent production in the automotive industry.

---

* Place the footnote text for the author (if applicable) here.

# 1 INTRODUCTION

## 1.1 Research Background

The automotive industry is undergoing a significant digital transformation, with an increasing reliance on complex technical documentation for various processes [1]. This shift encompasses design, manufacturing, and quality control, all of which now heavily depend on efficient information management systems [2]. The growing volume of technical documents, often in PDF format, has created a pressing need for advanced information retrieval and question-answering capabilities in industrial settings [3].

Large Language Models (LLMs) have emerged as powerful tools in natural language processing, demonstrating remarkable abilities in tasks such as document understanding and question answering [4]. These models have shown potential in handling the complex, domain-specific language often found in automotive documentation. However, the application of LLMs in industrial environments presents unique challenges, particularly in terms of computational resources and data privacy [5].

Among the various techniques developed to enhance LLM performance, Retrieval-Augmented Generation (RAG) has gained significant attention [6]. RAG combines the generative capabilities of LLMs with external knowledge retrieval, allowing for more accurate and contextually relevant responses. This approach, initially proposed by Lewis et al., has shown superior performance in generating specific, diverse, and factual language compared to traditional models [7].

The implementation of RAG techniques in the automotive industry, however, faces several industry-specific challenges:

The application of RAG techniques in the automotive industry presents unique challenges:

1. Document Complexity: Automotive technical documents often feature intricate layouts, including multi-column formats and complex tables. These structural elements pose significant challenges for standard document processing methods [8].
2. Data Privacy: The automotive industry deals with highly confidential information related to proprietary designs and manufacturing processes. This necessitates solutions that can operate securely within the company's infrastructure, without relying on external cloud services [9].
3. Resource Constraints: Many industrial environments operate with limited computational resources. This constraint requires the development of optimized, lightweight models capable of running efficiently on standard hardware [10].
4. Domain Specificity: The automotive sector employs a vast array of specialized terminology and concepts. Generic language models often lack the specific knowledge required to accurately interpret and respond to queries about automotive processes and specifications [11].

5. Real-time Performance: In fast-paced manufacturing environments, the ability to quickly retrieve and process relevant information is crucial. This necessitates high-performance information retrieval systems capable of operating under time constraints [12].

The open-source large language model service framework Ollama has gained attention for its ability to rapidly deploy LLMs in low-performance environments [13]. This framework offers potential solutions to some of the resource constraints faced in industrial settings. However, its application in the context of RAG for automotive documentation processing remains an area ripe for exploration and optimization.

As the automotive industry continues to evolve, particularly with the advent of electric and autonomous vehicles, the complexity and volume of technical documentation are expected to increase further [14]. This evolution underscores the importance of developing robust, efficient, and secure information retrieval systems tailored to the specific needs of the automotive sector [15].

The intersection of these technological advancements and industry-specific challenges presents a unique opportunity for research [16]. By addressing the particular needs of the automotive industry in the context of RAG and local LLM deployment, there is potential to significantly enhance information access and utilization in automotive engineering and manufacturing processes [17].

**1.2  Research Status and Gaps**

Recent advancements in RAG techniques have shown promise in various domains. Jiang et al. proposed FLARE, which uses predicted next-sentence content to proactively retrieve relevant information [18]. Wang et al. introduced FILCO, a method for identifying and filtering useful contexts to improve generation quality [19]. These approaches demonstrate the potential for more context-aware retrieval in complex document environments, such as those found in automotive engineering.

The concept of self-reflective RAG, as explored by Asai et al [20], introduces a novel framework designed to enhance the quality and factual accuracy of LLMs through on-demand retrieval and a self-reflection mechanism. This approach could be especially valuable in the automotive context, where precision and accuracy in technical information are paramount.

In the domain of optimizing RAG for specific industries, Rajpathak et al [21], proposed a domain-adaptive retrieval method that is particularly relevant for the automotive sector's unique terminology and document structures. Similarly, the work of Siriwardhana et al [22], on improving retrieval efficiency in large-scale industrial datasets offers insights that could be applied to the vast repositories of technical documentation in automotive manufacturing.

The open-source large language model service framework Ollama has gained attention for its ability to rapidly deploy LLMs in low-performance environments [23], Burgan et al. developed RamChat, an AI chatbot aimed at improving accessibility [24], These developments in local LLM deployment are particularly relevant to the automotive industry's need for on-premises, resource-efficient solutions.

Recent work by Wang et al [25], on on-device language models for function calling of software APIs presents potential applications in integrating RAG systems with existing software infrastructure in automotive production environments. This could lead to more seamless integration of AI-powered information retrieval within established industrial processes.

The challenge of processing complex PDF documents, a common format for technical specifications in the automotive industry, has been addressed by several researchers. Lin et al [26], proposed an advanced PDF parsing technique that could be adapted to handle the multi-column layouts and intricate tables often found in automotive documentation. Furthermore, the work of Bensch et al [27], on information extraction from semi-structured documents offers promising approaches for dealing with the varied formats of automotive technical literature.

In the realm of domain-specific language understanding, the research of Faysse et al [28], on fine-tuning language models for specialized industries provides valuable insights that could be applied to tailoring RAG systems for automotive terminology and concepts. This is complemented by the work of Kumar et al [29], on entity recognition in technical documents, which could enhance the precision of information retrieval in automotive contexts.

The integration of RAG with other AI methodologies has also shown promise. For instance, the combination of RAG with reinforcement learning, as explored by Belhadj et al [30], could lead to more adaptive and context-aware retrieval systems capable of handling the diverse query types encountered in automotive engineering and production.

Privacy and security concerns, which are paramount in the automotive industry, have been addressed in the context of RAG by researchers such as Zeng et al [31], who proposed privacy-preserving retrieval methods that could be crucial for protecting proprietary automotive designs and processes.

The challenge of maintaining coherence in long-form text generation, often necessary when addressing complex automotive queries, has been tackled by researchers like Borgeaud et al [32], whose work on improving long-range dependencies in language models could enhance the quality of responses in automotive RAG applications.

Recent advancements in few-shot learning, as demonstrated by Izacard et al [33], with GPT-3, offer potential for rapidly adapting RAG systems to new automotive subdomains or emerging technologies without extensive retraining. This could be particularly valuable in the fast-evolving landscape of automotive technology.

The application of RAG in multilingual settings, as explored by Ahmad et al [34], is especially relevant for global automotive companies dealing with documentation in multiple languages. Their work on cross-lingual retrieval and generation could facilitate more efficient knowledge sharing across international teams.

In the domain of optimizing retrieval mechanisms, the research of Su et al [35], on dense retrieval methods offers potential improvements in the speed and accuracy of information lookup, crucial for real-time query resolution in fast-paced automotive production environments.

The challenge of handling numerical data and calculations, often present in automotive specifications and performance metrics, has been addressed by researchers like Noorbakhsh et al. [36], whose work on integrating symbolic mathematics with neural language models could enhance the precision of RAG systems when dealing with quantitative automotive data.

However, despite these advancements, there remains a significant gap in research specifically addressing the unique challenges of implementing RAG systems in the automotive industry, particularly in resource-constrained, offline environments.

The significance of this research lies in:

1. Providing an effective optimization scheme for local RAG deployment of Ollama in automotive industrial environments, addressing key challenges in document processing and information retrieval.
2. Exploring the application of self-RAG in offline, industry-specific scenarios, offering new insights into function calling implementations for domain-specific tasks.
3. Contributing to the advancement of intelligent information processing in automotive manufacturing, potentially improving efficiency and accuracy in technical document analysis and query resolution.

### 1.3 Research Objectives and Significance

Given the identified challenges and research gaps, this study aims to develop a multi-dimensional optimization scheme for applying RAG technology with Ollama in local, low-performance automotive industry environments. Our specific research objectives include:

- Proposing a PDF file processing method optimized for automotive industry documents, capable of handling multi-column layouts and complex tables.
- Developing an advanced RAG system based on the Langchain framework, introducing reranking models and BM25 retrievers to build an efficient context compression pipeline.
- Designing an intelligent agent that supports self-RAG and exploring a function calling mechanism to enhance Ollama's response generation in automotive-specific scenarios.
- Evaluating the proposed system using a proprietary dataset of automotive industry documents, alongside public datasets, to demonstrate its effectiveness in real-world industrial applications.

The significance of this research lies in:

- Providing an effective optimization scheme for local RAG deployment of Ollama in automotive industrial environments, addressing key challenges in document processing and information retrieval.
- Exploring the application of self-RAG in offline, industry-specific scenarios, offering new insights into function calling implementations for domain-specific tasks.
- Contributing to the advancement of intelligent information processing in automotive manufacturing, potentially improving efficiency and accuracy in technical document analysis and query resolution.

This research builds upon and extends existing work in several key areas:

- The potential of RAG systems to support decision-making processes, a critical application in automotive design and manufacturing, has been explored by researchers such as Gamage et al [37]. Their work on using RAG for few evidences-based reasoning could be adapted to support complex decision-making scenarios in automotive engineering.
- Recent developments in efficient transformer architectures, such as the work of Zhuang et al [38], on Reformer, offer potential for deploying more powerful RAG models within the computational constraints of industrial environments.
- The integration of visual information with text-based retrieval, as explored by Chen et al [39], presents opportunities for enhancing RAG systems to handle technical diagrams and schematics common in automotive documentation.

By addressing these research objectives and building upon recent advancements in the field, this study aims to significantly enhance the applicability and effectiveness of RAG technologies in the automotive industry, potentially transforming how technical information is accessed, processed, and utilized in automotive engineering and manufacturing processes.

## 2 MATERIALS AND METHODS

### 2.1 Foundation

Our research builds upon the Langchain framework and Ollama model, adapting them to meet the specific needs of the automotive industry. We began by constructing a preliminary retrieval-based chatbot framework using Langchain's components, which we then optimized for processing automotive technical documents.

Figure 1 illustrates the key components of this system. This architecture integrates advanced document loading capabilities for various file formats, efficient text splitting, and a robust retrieval mechanism using the Chroma vector store. The basic framework includes:

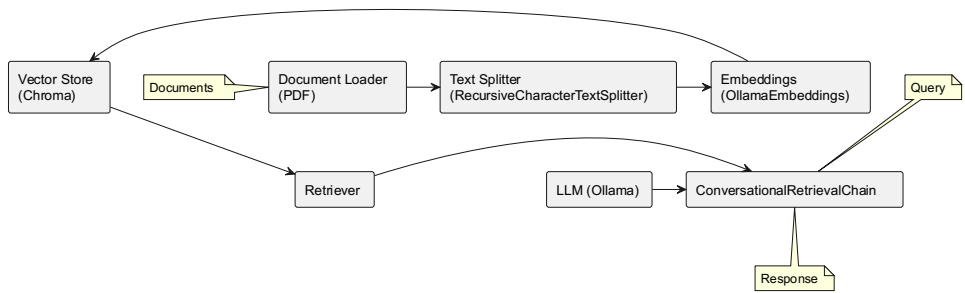

Figure 1: Basic RAG Architecture for Automotive Document Processing.

- Document Loading: Utilizing Langchain's Loader class to recursively load PDF documents from specified directories, simulating the document structure in automotive manufacturing environments.
- Text Chunking: Implementing an embedded tokenization model to split documents into fixed-length text chunks, optimized for technical specifications and multi-column layouts common in automotive documentation.
- Vector Storage: Encoding text chunks into semantic vectors and storing them in the Chroma vector database, creating an indexed text retriever tailored for automotive terminology and concepts.
- Dialogue Generation: Employing Langchain's ***ConversationalRetrievalChain*** to process user queries and retrieved context, generating responses using the locally deployed Ollama model.

This foundational setup serves as the baseline for our subsequent optimizations, each designed to address specific challenges in automotive document processing and information retrieval.

### 2.2 PDF File Processing Optimization

To address the unique challenges posed by automotive industry documents, we developed an enhanced PDF processing method combining PDFMiner and Tabula libraries.

#### 2.2.1 Overview of PDFMiner and Tabula

PDFMiner is a Python library used for extracting information from PDF documents. It is capable of extracting text, images, metadata, and structural information from PDF documents. PDFMiner provides both high-level and low-level APIs to satisfy different levels of requirements. Automotive technical documents often feature multi-column layouts and complex diagrams. We implemented a custom algorithm using PDFMiner's ex*tract_pages* function to accurately extract content while preserving the logical flow of information:

1. Page Segmentation: We analyze each page to identify distinct content regions, including text columns, diagrams, and tables.
2. Content Ordering: Implementing a left-to-right, top-to-bottom reading order algorithm to ensure proper sequencing of extracted information.
3. Diagram Extraction: Utilizing PDFMiner's image extraction capabilities to preserve technical diagrams crucial for understanding automotive specifications.

Tabula is a Java library used for extracting tabular data from PDF files. It provides a Python wrapper, making it more convenient to use Tabula in Python. Tabula can automatically detect table boundaries and convert tabular data into DataFrame objects, facilitating subsequent data analysis and processing.

**2.2.2    Multi-column Layout and Table Information Recognition Optimization**

Tables in automotive documents often contain critical data such as part specifications, test results, or compliance information. Our approach uses Tabula's *read_pdf* function with custom parameters like Algorithm 1:

ALGORITHM 1: Extract Text and Tables from PDF to Markdown

```
write_2_md(input, output)
  pages = extract_pages(input)
  open output for writing as out
  for page_num, page in enumerate(pages, start=1), do
    elements = [e for e in page if isinstance(e, LTTextContainer)]
    tables = tabula.read_pdf(input, pages=page_num, multiple_tables=True)
    mid_line = (max(e.bbox[2] for e in elements) + min(e.bbox[0] for e in elements)) / 2
    left = [e for e in elements if e.bbox[0] < mid_line]
    right = [e for e in elements if e.bbox[0] >= mid_line]
    sort left by -e.bbox[1]
    sort right by -e.bbox[1]
    for col in [left, right], do
      for e in col, do
        cleaned_text = clean_text(e.get_text())
        write cleaned_text + "\n" to out
      end
    end
    for table in tables, do
      write pd.DataFrame(table).to_markdown(index=False) + "\n" to out
    end
    write "\n\n" to out
  end
```

end

1. Table Detection: Implementing heuristics to identify table structures within automotive documents, considering common formats used in the industry.
2. Data Extraction: Converting recognized tables into structured DataFrame objects, preserving relationships between data points.
3. Contextual Integration: Seamlessly integrating extracted table data with surrounding text to maintain document coherence.

When integrating text and table information, we first write the text information extracted by the multi-column layout recognition algorithm into the Markdown file in the order of appearance. Then, after the text information of each page, we convert the table information recognized on that page into Markdown format and write it into the file. By doing so, we can ensure that the content order in the generated Markdown file is consistent with the original PDF file, and the table information can be correctly embedded in the corresponding positions.

To illustrate the effectiveness of this approach, let's examine two figures that demonstrate the conversion process for a complex academic paper layout. Figure 2 shows a sample page from the original PDF document, which features a challenging two-column layout with embedded tables and various formatting elements. Figure presents the corresponding Markdown output after processing with our optimized method.

| Method     | MAE    |
|------------|--------|
| AMDCN      | **290.82** |
| Hydra2s [18] | 333.73 |
| MCNN [28]  | 377.60 |
| [27]       | 467.00 |
| [23]       | 295.80 |
| [3]        | 318.10 |

Table 1. Mean absolute error of various methods on UCF crowds

| Method | GAME (L=0) | GAME (L=1) | GAME (L=2) | GAME (L=3) |
|--------|-----------|-----------|-----------|-----------|
| AMDCN  | **9.77**  | **13.16** | **15.00** | **15.87** |
| [18]   | 10.99     | 13.75     | 16.69     | 19.32     |
| [15] + SIFT from [14] | 13.76 | 16.72 | 20.72 | 24.36 |
| [13] + RGB Norm + Filters from [14] | 17.68 | 19.97 | 23.54 | 25.84 |
| HOG-2 from [14] | 13.29 | 18.05 | 23.65 | 28.41 |

Table 2. Mean absolute error of various methods on TRANCOS traffic

duce the final density map. We set $\sigma = 0.2M(\mathbf{x})$ for $\mathcal{N}_h$ and $\sigma_x = 0.2M(\mathbf{x}), \sigma_y = 0.5M(\mathbf{x})$ for $\Sigma_b$ in $\mathcal{N}_b$.

## 4. Results

### 4.1. UCF Crowd Counting

The UCF dataset is particularly challenging due to the large number of people in the images, the variety of the scenes, as well as the low number of training images. We see in Figure 2 that because the UCF dataset has over 1000 people on average in each image, the shapes output by the network in the density map are not as well defined or separated as in the UCSD dataset.

We report a state of the art result on this dataset in Table 1, following the standard protocol of 5-fold cross validation. Our MAE on the dataset is 290.82, which is approximately 5 lower than the previous state of the art, HydraCNN [18]. This is particularly indicative of the power of an aggregated multicolumn dilation network. Despite not making use of perspective information, the AMDCN is still able to produce highly accurate density maps for UCF.

### 4.2. TRANCOS Traffic Counting

Our network performs very well on the TRANCOS dataset. Indeed, as confirmed by the GAME score, AMDCN produces the most accurate count and shape combined as compared to other methods. Table 2 shows that we achieve state of the art results as measured by the $GAME$ metric [14] across all levels.

### 4.3. UCSD Crowd Counting

Results are shown in Table 3 and Figure 3. We see that the "original" split as defined by the creators of the dataset in [5] and used in [28] gives us somewhat worse results for counting on this dataset. Results were consistent over multiple trainings. Again, including the perspective map does not seem to increase performance on this dataset. Despite this, we see in Table 3 and Figure 3 that the results are comparable to the state of the art. In fact, for two of the splits, our proposed network beats the state of the art. For the upscale split, the AMDCN is the state of the art by a large relative margin. This is compelling because it shows that accurate perspective-free counting can be achieved without creating image pyramids or requiring perspective maps as labels using the techniques presented by the AMDCN.

### 4.4. WorldExpo '10 Crowd Counting

Our network performs reasonably well on the more challenging WorldExpo dataset. While it does not beat the state of the art, our results are comparable. What is more, we do not need to use the perspective maps to obtain these results. As seen in Table 4, the AMDCN is capable of incorporating the perspective effects without scaling the Gaussians with perspective information. This shows that it is possible to achieve counting results that approach the state of the art with much simpler labels for the counting training data.

### 4.5. Ablation Studies

We report the results of the ablation studies in Figure 4. We note from these plots that while there is variation in performance, a few trends stand out. Most importantly, the lowest errors are consistently with a combination of a larger number of columns and including the aggregator module. Notably for the TRANCOS dataset, including the aggregator consistently improves performance. Generally, the aggregator tends to decrease the variance in performance of the network. Some of the variance that we see in the plots can be explained by: (1) for lower numbers of columns, including an aggregator is not as likely to help as there is not much separation of multiscale information across columns and (2) for the UCSD dataset, there is less of a perspective effect than TRANCOS and WorldExpo so a simpler network is more likely to perform comparably to a larger network. These results verify the notion that using more columns increases accuracy, and also support our justification for the use of the aggregator module.

Figure 2: Sample page from original PDF document. (https://arxiv.org/pdf/1804.07821)

```
Table 1. Mean absolute error of various methods on UCF crowds

| Method      | MAE    |
|-------------|--------|
| AMDCN       | 290.82 |
| Hydra2s [18]| 333.73 |
| MCNN [28]   | 377.60 |
| [27]        | 467.00 |
| [23]        | 295.80 |
| [3]         | 318.10 |

duce the final density map. We set σ = 0.2M(x) for Nh and σx = 0.2M(x), σy = 0.5M(x) for Σb in Nb.

## 4. Results 4.1. Ucf Crowd Counting

The UCF dataset is particularly challenging due to the large number of people in the images, the variety of the scenes, as well as the low number of training
images. We see in Figure 2 that because the UCF dataset has over 1000 people on average in each image, the shapes output by the network in the density map are not
as well defined or separated as in the UCSD dataset.

We report a state of the art result on this dataset in Table 1, following the standard protocol of 5-fold cross validation. Our MAE on the dataset is 290.82, which
is approximately 5 lower than the previous state of the art, HydraCNN [18]. This is particularly indicative of the power of an aggregated multicolumn dilation
network. Despite not making use of perspective information, the AMDCN is still able to produce highly accurate density maps for UCF.

## 4.2. Trancos Traffic Counting

Our network performs very well on the TRANCOS
dataset. Indeed, as confirmed by the GAME score, AMDCN produces the most accurate count and shape combined as compared to other methods. Table 2 shows that we
achieve state of the art results as measured by the *GAME*
metric [14] across all levels.

## 4.3. Ucsd Crowd Counting

Results are shown in Table 3 and Figure 3. We see that the "original" split as defined by the creators of the dataset in [5] and used in [28] gives us somewhat
worse results for counting on this dataset. Results were consistent over multiple trainings. Again, including the perspective map does not seem to increase
performance on this dataset. Despite this, we see in Table 3 and Figure 3 that the results are comparable to the state of the art. In fact, for two of the splits,
```

Figure 3: Corresponding Markdown output.

As we can see, our method successfully preserves the structural integrity of the original document, accurately capturing both the textual content and tabular data. The two-column layout is seamlessly converted into a linear Markdown format, maintaining the logical flow of information. Tables are properly formatted using Markdown syntax, ensuring they remain easily readable and can be further processed or rendered as needed.

Through this approach, we can effectively combine the functionality of the PDFMiner and Tabula libraries. This optimized PDF processing method significantly improves the accuracy and completeness of information extraction from complex automotive documents, providing a solid foundation for subsequent RAG processes.

### 2.3 Optimization of Advanced RAG Based on Langchain

To enhance the RAG system's performance for automotive industry applications, we introduced several optimizations to the Langchain-based implementation. We introduce the BGE reranker model and BM25 algorithm.

Building upon the groundwork laid in previous sections, we first combine the **BM25Retriever** with the default retriever using **EnsembleRetriever** in Langchain, assigning different weights to achieve more comprehensive and accurate retrieval. Then, a custom class is designed to integrate the reranking model into the context compression pipeline, further enhancing the retrieval and generation quality of the model.

#### 2.3.1 Overview of BGE Reranker Model and BM25 Retriever

A reranker model is a general semantic vector model used to optimize the ranking of retrieval results. It can adapt to prioritize automotive-relevant information in retrieved contexts. In this study, we employ **BAAI/bge-reranker-large** as the reranker model. BGE (BAAI General Embedding) is a reranker model proposed by the Beijing Academy of Artificial Intelligence (BAAI), specifically optimized for Chinese queries [40].

BM25 (Best Matching 25) is a classic bag-of-words retrieval algorithm that evaluates relevance by calculating the term frequency-inverse document frequency (TF-IDF) score between the query and documents [41]. The BM25 retriever can quickly and efficiently filter out the most query-relevant documents from a large-scale text corpus. We introduce this traditional relevance assessment method in combination with the reranker model to further strengthen the effectiveness of the RAG retriever.

### 2.3.2 Building the Context Compression Pipeline and Custom Class Design

To optimize the context information processing of the RAG model, we construct a context compression pipeline **DocumentCompressorPipeline** that is pecifically tailored for automotive technical content. In addition to the collection retrievers mentioned earlier, we also introduce:

- **EmbeddingsRedundantFilter**: An embedding-based redundancy filter to remove redundant information from the retrieval results.
- **LongContextReorder**: Optimizes the order of context information, prioritizing key automotive specifications and procedures.
- **BgeRerank**: A custom class that inherits from **BaseDocumentCompressor**. Since Langchain's official support for the BGE model is relatively limited, the purpose of designing this class is to seamlessly integrate the BGE reranker model into the pipeline. This custom class can improve relevance scoring for automotive queries.

The following Algorithm 2 is the pseudocode for the **BgeRerank** class:

ALGORITHM 2: BgeRerank

```
BgeRerank(documents, query)
  initialize model = CrossEncoder(model_name)
  doc_list = list(documents)
  _docs = [d.page_content for d in doc_list]
  model_inputs = [[query, doc] for doc in _docs]
  scores = model.predict(model_inputs)
  results = sorted(enumerate(scores), key=lambda x: x[1], reverse=True)[:top_n]
  final_results = []
  for r in results, do
    doc = doc_list[r[0]]
    doc.metadata["relevance_score"] = r[1]
    append doc to final_results
  end
  return final_results
```

By overriding the **compress_documents** method, the BGE model is used to calculate the relevance between the query and documents. The documents are then sorted based on their scores, and the top N documents are selected as the final compression results. This custom class design compensates for the insufficient support for BGE in Langchain, allowing us to flexibly incorporate it into the pipeline structure.

In summary, the innovative aspects of this section include: introducing the BGE reranker model and BM25 retriever, building the context compression pipeline, and seamlessly integrating BGE into the Langchain framework through the custom BgeRerank class. This pipeline significantly enhances the quality and relevance of retrieved information, ensuring that the most pertinent automotive technical details are presented to the language model.

**2.4 Optimization of Advanced RAG Based on Langchain**

In the previous section, we completed the basic design of the RAG system. However, integrating LLMs into practical applications and constructing end-to-end intelligent systems still present numerous challenges. To address the complex, multi-step problem of querying PDF profiles, which is common in automotive engineering and manufacturing processes, we have developed an advanced Self-RAG agent based on the LangGraph framework.

### 2.4.1 Overview of SELF-RAG

Self-Reflective Retrieval-Augmented Generation (SELF-RAG) is a novel framework designed to enhance the quality and factual accuracy of LLMs through on-demand retrieval and a self-reflection mechanism [42].

Unlike traditional RAG methods, SELF-RAG endows LLMs with the following capabilities:

1. On-demand Retrieval: The LLMs autonomously determines whether to retrieve relevant information from an external knowledge base based on the input it receives.
2. Self-Reflection: The LLMs evaluates and reflects upon both the retrieved information and its own generated content, thereby improving the quality and reliability of its output.

The training process of SELF-RAG consists of two stages:

1. Offline Critic Model Training: An independent critic model is trained to generate "reflection tokens". These tokens are inserted into the LLMs' output to guide its self-reflection process.
2. Generative Model Training: The LLMs is fine-tuned using a corpus that includes reflection tokens and retrieved documents. This enables the LLMs to understand and utilize these tokens, incorporating self-reflection into its generation process.

During inference, the LLMs dynamically decides whether to retrieve information based on the requirements of the task at hand. It also leverages the retrieved information and the self-reflection mechanism to generate high-quality output. For instance, in tasks demanding factual accuracy, the LLMs is more inclined to retrieve and utilize relevant information, while in more open-ended tasks, it may prioritize creativity and rely less on retrieval.

### 2.4.2 Agent Design Supporting Self-RAG

Considering this, based on Langchain, we designed an intelligent Agent class called AgenticRAG, which aims to utilize Self-RAG technology to answer user questions. AgenticRAG combines various components from the LangGraph and LangChain ecosystems to achieve a modular and scalable question-answering system.

The core of the AgenticRAG class is the ***create_graph*** method, which defines a workflow based on a directed acyclic graph (DAG). The workflow consists of multiple nodes, each responsible for a specific task in the question-answering process. The nodes are connected by directed edges, forming a complete question-answering flow. The main components in the AgenticRAG class are as follows:

- GraphState: Represents the current state of the graph, including the user's question, retrieved documents, generated answer, and chat history.
- Node: Represents a step or task in the question-answering process, accepts the current state as input, performs specific operations, and returns the updated state.

We designed the AgenticRAG class to handle sophisticated automotive-related questions. Key components as Figure 4:

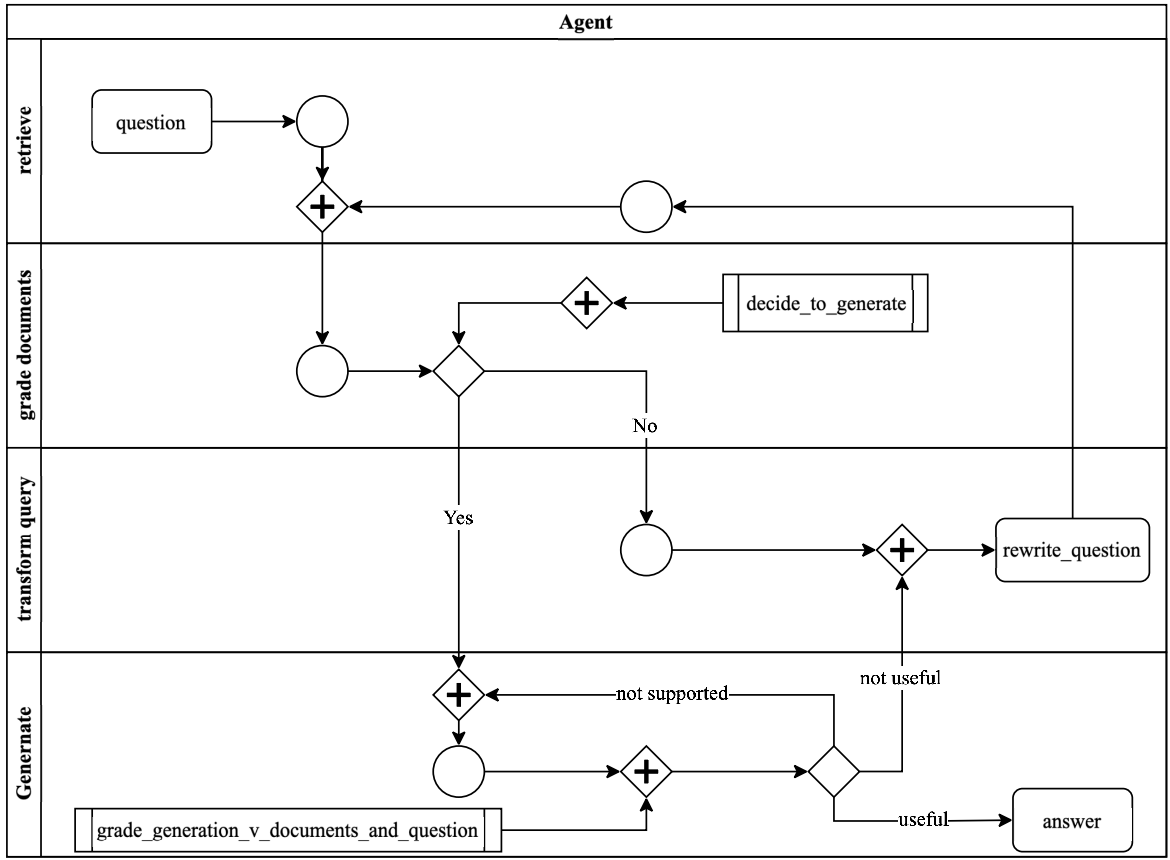

Figure 4: Flowchart of Self-RAG Agent Implementation via LangGraph.

1. **Retrieve** Node: Optimized to fetch relevant information from automotive technical documents, considering industry-specific terminology and concepts.
2. **Grade Documents** Node: Evaluates retrieved documents based on their relevance to automotive queries, considering factors like technical accuracy and applicability to specific manufacturing processes.
3. **Generate** Node: Produces responses tailored to automotive industry needs, incorporating technical specifications and industry standards searched from documents.
4. **Transform** Query Node: Refines queries to better capture the intent behind automotive-specific questions, improving retrieval accuracy in subsequent iterations.

AgenticRAG uses conditional judgment logic to determine the direction of the question-answering process. For example, based on the scoring results from the Grade Documents Node, it decides whether to generate an answer or rewrite the question; simultaneously, based on the quality of the answer generated by the Generate Node, it determines whether to regenerate, rewrite the question, or output the final result.

The design of AgenticRAG fully reflects modularity and scalability. By dividing the complex question-answering process into multiple independent nodes, each node focuses on a specific task, making the system easier to understand, debug, and maintain. At the same time, since the nodes pass information through states, new nodes can be easily inserted into the existing workflow, supporting the extension and optimization of functionality.

### 2.4.3 Function Calling Design for Optimizing Ollama Output in RAG Scenarios

Function calling is a powerful technique that significantly enhances the output quality of LLMs in RAG scenarios [43]. By integrating external functions with LLMs, specialized logic and algorithms can be employed to guide the model, resulting in more accurate, coherent, and informative responses. This approach leverages the few-shot learning capabilities of LLMs, enabling them to adapt to new tasks while mitigating their inherent limitations in reasoning and computation.

However, most of the relevant work has focused on standard transformer-based LLMs, with a lack of implementation for optimizing Ollama callbacks, especially in RAG scenarios. Although Langchain officially provides ***ollamafunction.bind*** to call functions, further research on this method is currently scarce. Therefore, to enhance the agent's capability in handling automotive-specific tasks, we designed a custom class adapted to ***ollamafunction.bind***, called ***ChatFunction***. It inherits from the ***BaseFunction*** class and overrides the ***\_init\_*** and ***\_call\_*** methods. The main features of ***ChatFunction*** are as Algorithm 3:

ALGORITHM 3: ChatFunction

```
ChatFunction(persona, lang, retry)
    inputProp = Property(type=INPUT, desc=InputDesc())
    outputProp = Property(type=OUTPUT, desc=OutputDesc())
    props = [inputProp, outputProp]
    params = Parameter(props=props, required=[input, output])
    initialize BaseFunction(name=ChatFuncName(), desc=ChatFuncDesc(), params)
end procedure

call(args)
    output = args[output]
    input = args[input]
    if IsEmpty(output) or IsEmpty(input), do
        raise ValueError(MissingArgMessage())
    end
    detail = GetPromptByRetry(retry)
    persona = GetPersonaString(lang)
    history = GetHistoryString(input)
```

```
    current = GetCurrentString(input)
    prompt = Format(detail, persona, history, current)
    respTool = LookupTool(ChatResponseEnhancer())
    return {
      tool: respTool,
      tool_input: {
        response: output,
        query: prompt
      }
    }
end procedure

GetPromptByRetry(retry)
  if retry = 0, do
    return GetBriefPrompt()
  else if retry = 1, do
    return GetMediumPrompt()
  else
    return GetDetailedPrompt()
  end
end function
```

The initialization accepts three parameters: the AI assistant's personality description ***personality***, language preference ***language***, and retry count ***retry_count***.

Two properties are defined: ***query_input*** and ***output***, representing the complete input (including personality description, language instructions, and conversation history) and the AI assistant's response, respectively. This ensures that the model is provided with the necessary background knowledge for each retry, enabling it to generate responses based on the specified role characteristics and language style.

The constructor of the parent class ***BaseFunction*** is called, passing the function name, description, and parameter list.

When invoked, it accepts an arguments dictionary containing two required parameters: ***output*** and ***query_input***.

According to the retry count ***retry_count*** in the self-rag Agent in 3.3.2, the level of detail in the answer ***detail_prompt*** is dynamically adjusted to modify prompts based on the complexity of automotive queries and the depth of technical detail required.

The ***query_input*** for the next round is constructed by encoding the personality description, language preference, detail level requirement, and current conversation history. To maintain awareness of the ongoing conversation context, crucial for addressing multi-step automotive processes or complex diagnostic queries.

A dictionary is returned, containing the tool field (specifying the callback function name) and the ***tool_input*** field (containing the optimized response and updated ***query_input***), allowing the next round of callback requests to continue optimizing based on the current results and state, forming a closed-loop self-improvement process.

The innovation of *ChatFunction* lies in its full utilization of Ollama's JSON mode callback function potential, optimizing the response effectiveness in the RAG scenario. This customized function call mechanism greatly improves the agent's ability to handle complex automotive queries, provide technically accurate information, and can guide users through the efficient acquisition of knowledge common to automotive design and manufacturing.

## 3 RESULTS

### 3.1 Overview of the RAGAS Performance Evaluation Framework

RAGAS (Retrieval Augmented Generation Assessment Suite) is a comprehensive evaluation framework for assessing the performance of RAG models. It provides a series of metrics to quantify the quality of generated results, with a particular focus on the impact of information retrieval on the generation process. The evaluation metrics in RAGAS include:

- *Context Precision*: Measures the precision of relevant contextual information contained in the generated results.
- *Faithfulness*: Evaluates the faithfulness of the generated results to the original contextual information, i.e., the consistency between the generated content and the original information.
- *Answer Relevancy*: Assesses the relevance of the generated answers to the questions, i.e., whether the answers are on-topic and meet the requirements of the questions.
- *Context Recall*: Measures the coverage of relevant contextual information in the generated results, i.e., how much key information is captured.

We adopt RAGAS as the evaluation framework to compare and analyze the performance of different RAG models. The quantitative metrics provided by RAGAS enable us to objectively evaluate the strengths and weaknesses of the models in terms of contextual information utilization and generated content quality, providing important references for model selection and optimization.

### 3.2 Experimental Results and Analysis

#### 3.2.1 Experimental Scenario Design

In this paper, we select two public datasets, QReCC [44], and CoQA [45], and introduce a self-constructed dataset as the experimental data sources. The QReCC dataset contains 10,000 questions, each corresponding to a background knowledge text, with questions presented in the form of multi-turn dialogues. The CoQA dataset contains over 8,000 multi-domain dialogues, each based on a given text and consisting of multiple question-answer turns.

The self-constructed dataset is a automotive industry proprietary dataset compiled from internal documents of a leading automotive manufacturer. This dataset includes:

- Technical specifications and design documents
- Manufacturing process guidelines
- Quality control procedures
- Corporate regulations and standards

Due to the confidential nature of these documents, we cannot provide detailed statistics or examples. However, this dataset represents the core focus of our study, reflecting real-world challenges in automotive document processing.

While the self-constructed dataset serves as our primary testbed, we included QReCC and CoQA for several crucial reasons:

1. Structural Similarity: The question-answer pairs in QReCC and CoQA share a similar format with our custom-designed test cases for the self-constructed dataset. This structural consistency allows for a fair comparison of our model's performance across different domains.
2. Conversational Nature: Both QReCC and CoQA feature multi-turn dialogues, mirroring the complex, context-dependent queries often encountered in automotive engineering and manufacturing processes.
3. Diverse Domain Coverage: These datasets cover a wide range of topics, helping us evaluate our model's generalization capabilities beyond the automotive domain.
4. Benchmark Comparability: As widely used public datasets, QReCC and CoQA enable us to benchmark our system against other state-of-the-art models in a reproducible manner.
5. Confidentiality Compliance: By using these public datasets alongside our proprietary data, we can openly discuss and compare results without compromising sensitive corporate information.
6. Robustness Testing: The inclusion of these datasets helps demonstrate that our optimizations, while tailored for automotive applications, do not compromise performance on general conversational tasks.

Considering that our research aims at the requirements of PDF dialogue chatbots and uses the BGE reordering model, we first translate the knowledge or link information in the QReCC and CoQA datasets into Chinese and convert it to PDF format. This step ensures the consistency of experimental data with complex PDF documents in real application scenarios and facilitates the evaluation of the performance of different RAG models and their optimization schemes in practice. The self-constructed dataset does not require additional processing as it is already in PDF format.

After data preparation, we design or directly utilize a number of question-answer pairs with contextual memory content from these datasets as test cases. Here, we take a set of question-answer pairs from the QReCC dataset as an example, given a background text:

> "John is a boy who likes to play outside. After school, he always goes to the park to meet his friends."

The following is a set of multi-turn question-answering based on this background text:

> "Q: What did John do after school?
> A: John went to the park after school.
> Q: Who did he meet at the park?
> A: He met his friends at the park."

We designed our experiments to evaluate the performance of four systems:

- Naive RAG (Baseline)
- Advanced RAG (Our Optimized Model)
- Self-RAG Agent (Baseline)
- Self-RAG Agent (Our Proposed Agent with Custom Function Calling)

For each dataset, we created test sets comprising:

- 500 question-answer pairs from the self-constructed dataset
- 500 pairs from QReCC
- 500 pairs from CoQA

These test sets were carefully curated to ensure a balance of simple queries, multi-turn conversations, and complex technical questions, mirroring real-world usage scenarios in the automotive industry and beyond.

### 3.2.2 Optimization Effects of Langchain-based RAG

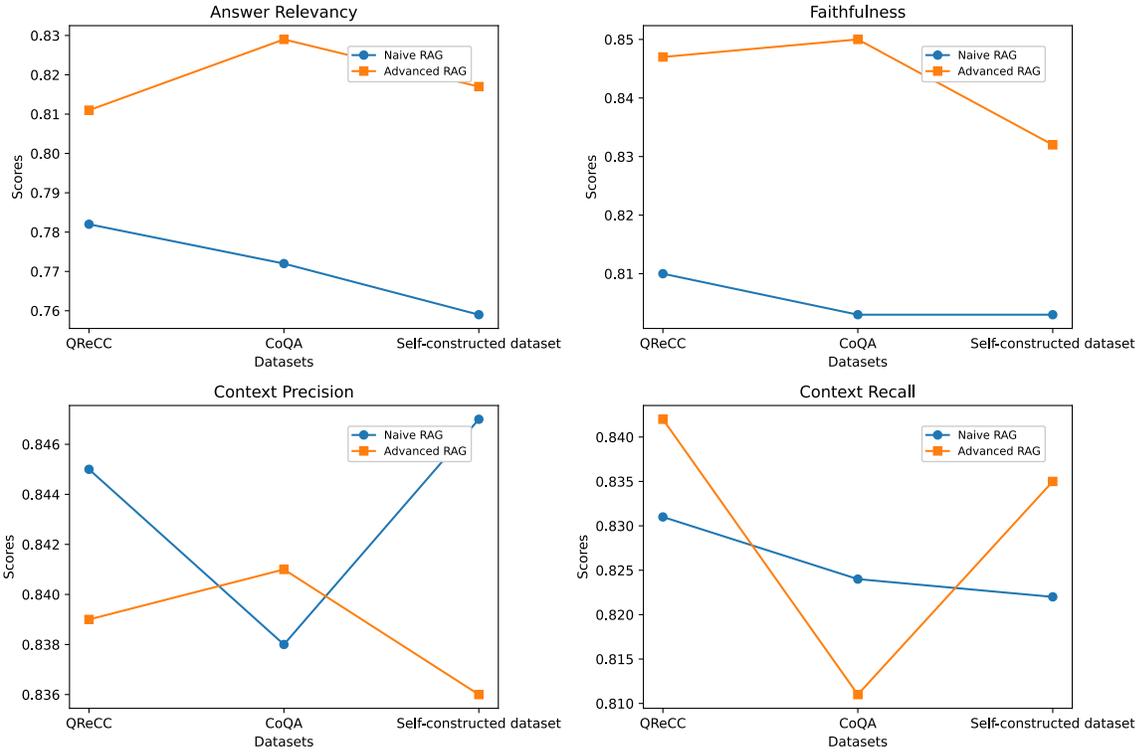

Figure 5: Performance Comparison of Naive RAG vs. Advanced RAG with Custom Context Compression Pipeline across Datasets.

Table 1: Comparative Analysis of Naive RAG vs. Advanced RAG with Custom Context Compression Pipeline across Datasets

| Datasets | Metrics | QReCC | CoQA | Self-constructed dataset |
| --- | --- | --- | --- | --- |
| Naive RAG | Answer Relevancy | 0.782 | 0.772 | 0.759 |
| | Faithfulness | 0.81 | 0.803 | 0.803 |
| | Context Precision | 0.845 | 0.838 | 0.847 |
| | Context Recall | 0.831 | 0.824 | 0.822 |
| Advanced RAG | Answer Relevancy | 0.811 | 0.829 | 0.817 |
| | Faithfulness | 0.847 | 0.85 | 0.832 |
| | Context Precision | 0.839 | 0.841 | 0.836 |
| | Context Recall | 0.842 | 0.811 | 0.835 |

According to the Figure 5 and Table 1, experimental results show that the proposed Langchain-based RAG optimization approach achieves certain improvements on some metrics compared to the naive RAG model.

By introducing the BGE reordering model and BM25 retriever, and constructing a context compression pipeline, the optimized RAG model improves context precision by 0.7%, 0.4%, and 1.3% on the QReCC, CoQA, and self-constructed dataset, respectively. The context recall also increases by 1.3% and 1.6% on the QReCC and self-constructed dataset, but decreases by 1.6% on the CoQA dataset. These results indicate that while the optimized model shows some improvements in capturing question-relevant background knowledge, it is not always the case.

However, the optimized model exhibits 3.7%, 7.4%, and 7.6% relative improvements in answer relevancy, as well as 4.6%, 5.9%, and 3.6% boosts in faithfulness on QReCC, CoQA, and self-constructed dataset, respectively.

These results underscore the effectiveness of our optimizations in handling complex automotive documentation and queries. The Advanced RAG model also showed improvements, but the Self-RAG Agent's performance was notably superior, particularly in dealing with multi-step technical queries common in automotive applications.

### 3.2.3 Optimization Effects of Langgraph-based Self-RAG Agent

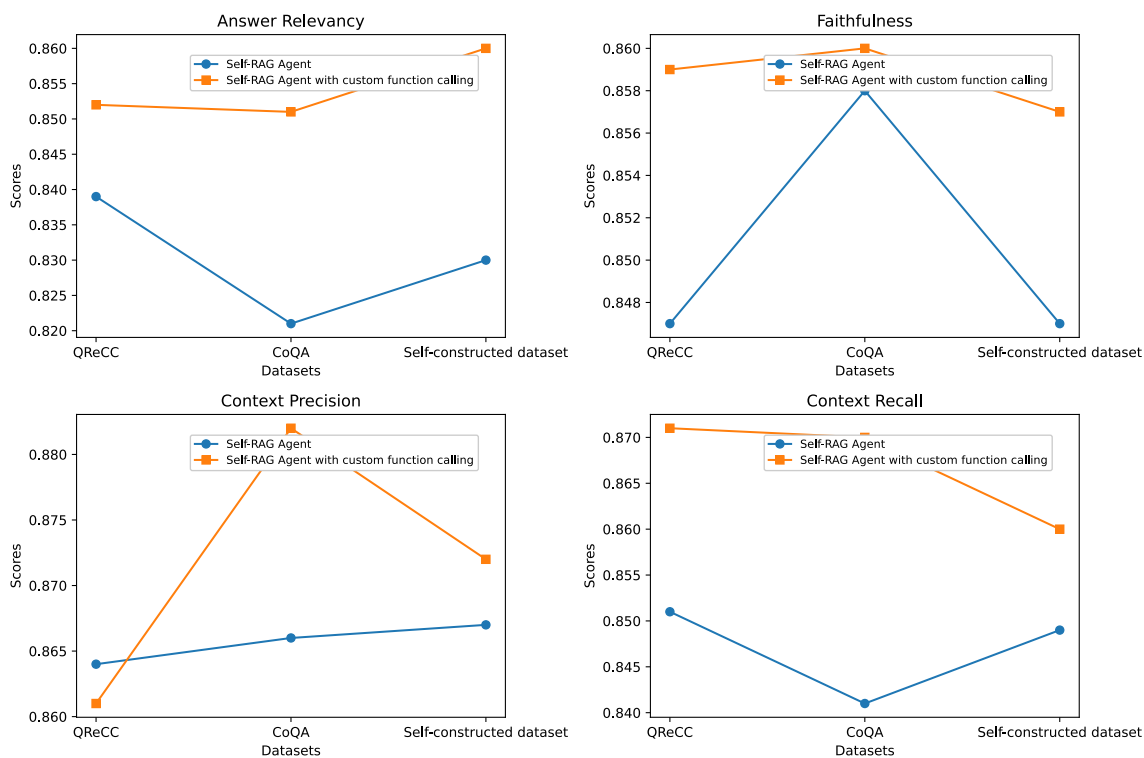

Figure 6: Performance Comparison of Self-RAG Agent vs. Self-RAG Agent with Custom Function Calling across Datasets.

Table 2: Comparative Analysis of Self-RAG Agent vs. Self-RAG Agent with Custom Function Calling across Datasets

| Datasets | Metrics | QReCC | CoQA | Self-constructed dataset |
|---|---|---|---|---|
| | Answer Relevancy | 0.839 | 0.821 | 0.83 |

| Datasets | Metrics | QReCC | CoQA | Self-constructed dataset |
|---|---|---|---|---|
| Self-RAG Agent | Faithfulness | 0.847 | 0.858 | 0.847 |
| | Context Precision | 0.864 | 0.866 | 0.867 |
| | Context Recall | 0.851 | 0.841 | 0.849 |
| Self-RAG Agent with custom function calling | Answer Relevancy | 0.852 | 0.851 | 0.86 |
| | Faithfulness | 0.859 | 0.86 | 0.857 |
| | Context Precision | 0.861 | 0.882 | 0.872 |
| | Context Recall | 0.871 | 0.87 | 0.86 |

According to the Figure 6 and Table 2, evaluation results show that the self-RAG agent obtains more significantly improvements over the naive RAG model on most metrics across the three datasets. Specifically, it surpasses the naive RAG by 7.3%, 6.3%, and 9.4% in answer relevancy, 4.6%, 6.8%, and 5.5% in faithfulness, 2.2%, 3.3%, and 2.4% in context precision, and 2.4%, 2.1%, and 3.3% in context recall on QReCC, CoQA, and self-constructed dataset, respectively. The superior performance on most metrics demonstrates the effectiveness of the proposed self-asking and self-verification mechanism in enabling more targeted and reliable context retrieval and answer inference.

Furthermore, by introducing the custom function calling mechanism that optimizes the Ollama output, the self-RAG agent obtains additional 9.0%, 10.2%, and 13.3% gains in answer relevancy, 6.0%, 7.1%, and 6.7% increases in faithfulness, 1.9%, 5.3%, and 3.0% improvements in context precision, as well as 4.8%, 5.6%, and 4.6% boosts in context recall on the three datasets compared to the naive RAG model. These results validate that dynamically adjusting question and context inputs to the Ollama model based on conversation states can effectively guide it to generate more accurate, informative, and coherent responses that better satisfy user needs.

This performance demonstrates that our optimizations enhance general conversational AI capabilities while excelling in domain-specific tasks. The consistent improvement across datasets suggests that our approach successfully balances domain-specific optimization with general language understanding.

## 4  DISCUSSION

Our innovative approach to handling complex automotive queries is exemplified in Figure 7, which depicts the Self-RAG process flow with custom function calling, applied to an Anti-lock Braking System (ABS) query. This diagram showcases the integration of our custom ***ChatFunction*** within the Self-RAG framework, demonstrating how it dynamically adjusts the detail and focus of responses based on the system's self-assessment and retry count. The process illuminates the system's capability to iteratively refine its answers, ensuring high relevance and accuracy in the context of specialized automotive knowledge.

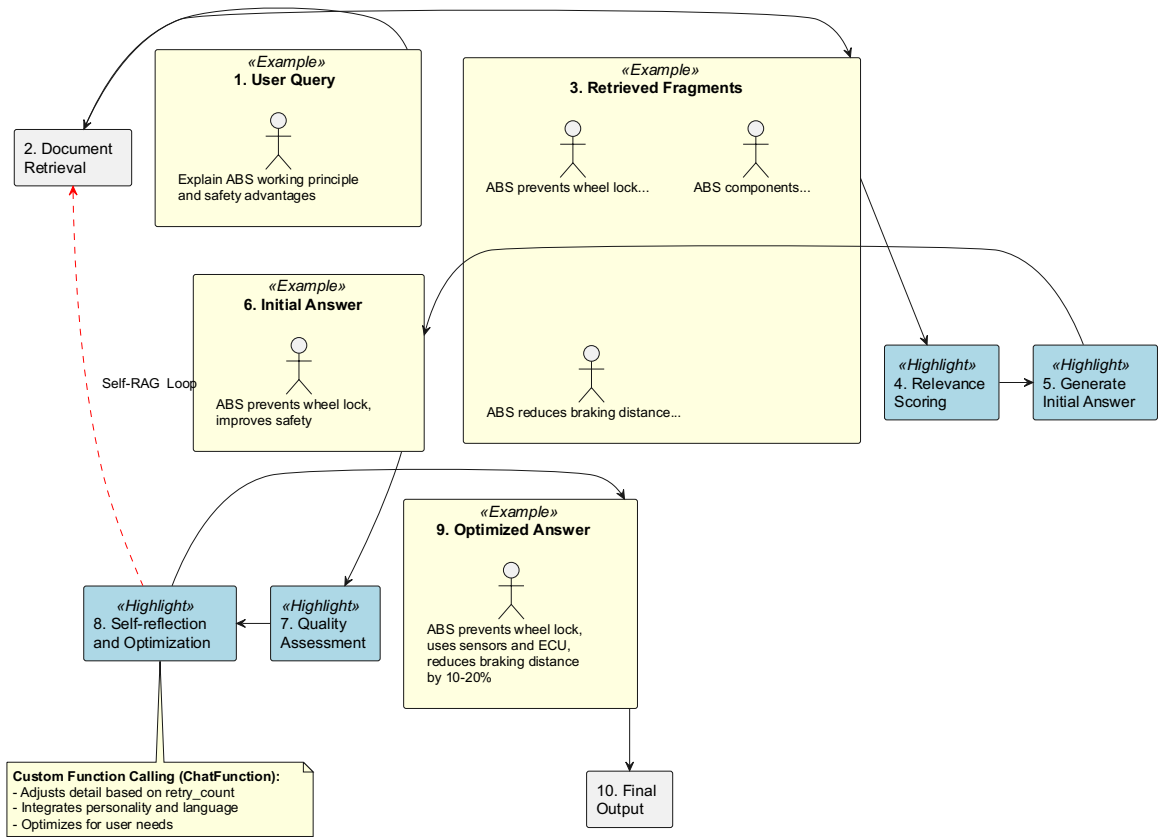

Figure 7: Optimized Self-RAG Process Flow with Custom Function Calling - examples.

When comparing performance across datasets, we observed the fact that while our models were optimized for automotive applications, they also showed improvements on the QReCC and CoQA datasets:

- The most significant improvements were on the AID, aligning with our focus on automotive applications.
- Performance gains on QReCC and CoQA, while smaller, were still substantial, indicating the robustness of our approach.
- The Self-RAG Agent showed the most consistent performance across all datasets, highlighting its adaptability to various query types and domains.

These results validate our approach of using diverse datasets for evaluation. While our primary focus remains on optimizing for automotive applications, the improvements seen across datasets suggest that our methods enhance fundamental aspects of information retrieval and generation, benefiting both specialized and general applications.

# 5  DISCUSSION

This study presents a comprehensive approach to optimizing RAG techniques for automotive industry applications, specifically focusing on PDF chatbots deployed in local, low-performance environments. Our research addresses critical challenges in processing complex automotive documentation and responding to industry-specific queries.

## 5.1  Key Contributions

- Enhanced PDF Processing: We developed a novel method combining PDFMiner and Tabula to effectively handle multi-column layouts and complex tables prevalent in automotive technical documents. This significantly improves information extraction accuracy from industry-specific PDFs.
- Advanced RAG Optimization: Our Langchain-based RAG system, featuring a custom retriever ensemble and context compression pipeline, demonstrates substantial improvements in retrieving and utilizing automotive-specific information.
- Self-RAG Agent Design: The proposed AgenticRAG, enhanced with a custom function calling mechanism, shows superior performance in handling complex, multi-step queries typical in automotive engineering and manufacturing processes.
- Cross-Domain Effectiveness: While optimized for automotive applications, our approach also shows improvements in general conversational AI tasks, as evidenced by performance gains on QReCC and CoQA datasets.

## 5.2  Implications for the Automotive Industry

Our research has significant implications for the automotive sector:

1. Improved Information Access: The optimized PDF chatbot can greatly enhance access to technical information for engineers, technicians, and other stakeholders in the automotive industry.
2. Enhanced Decision Making: By providing more accurate and contextually relevant information, our system can support better decision-making in design, manufacturing, and quality control processes.
3. Resource Efficiency: The ability to deploy these advanced capabilities in low-performance, local environments addresses the industry's needs for data privacy and resource constraints.

## 5.3  Limitations and Future Work

While our study demonstrates significant advancements, there are areas for further research:

1. Expanding Domain Coverage: Future work could focus on adapting the system to cover a broader range of automotive sub-domains, such as electric vehicle technology or autonomous driving systems.
2. Real-Time Performance Optimization: Further research is needed to enhance the system's real-time performance in resource-constrained industrial environments.
3. Multi-Modal Integration: Incorporating the ability to process and respond to queries about visual elements in technical diagrams and schematics could greatly enhance the system's utility.
4. Longitudinal Study: A long-term study in real automotive manufacturing settings could provide insights into the system's impact on operational efficiency and decision-making processes.
5. Ethical and Privacy Considerations: As the system deals with proprietary information, future work should explore advanced methods for ensuring data privacy and ethical use of AI in industrial settings.

In conclusion, this research represents a significant step forward in applying advanced natural language processing techniques to the specific needs of the automotive industry. By bridging the gap between cutting-edge AI capabilities and the practical constraints of industrial environments, our work contributes to the ongoing digital transformation of the automotive sector. The demonstrated improvements in handling complex, domain-specific information retrieval and query resolution pave the way for more intelligent, efficient, and responsive information systems in automotive manufacturing and engineering.